\documentclass[12pt,aip,preprint]{revtex4-1}
\usepackage[latin9]{inputenc}
\usepackage{array}
\usepackage{refstyle}
\usepackage{mathrsfs}
\usepackage{amsmath}
\PassOptionsToPackage{version=3}{mhchem}
\usepackage{mhchem}

\makeatletter

\AtBeginDocument{\providecommand\figref[1]{\ref{fig:#1}}}
\AtBeginDocument{\providecommand\tabref[1]{\ref{tab:#1}}}
\providecommand{\tabularnewline}{\\}
\RS@ifundefined{subsecref}
  {\newref{subsec}{name = \RSsectxt}}
  {}
\RS@ifundefined{thmref}
  {\def\RSthmtxt{theorem~}\newref{thm}{name = \RSthmtxt}}
  {}
\RS@ifundefined{lemref}
  {\def\RSlemtxt{lemma~}\newref{lem}{name = \RSlemtxt}}
  {}

\usepackage{siunitx}
\usepackage{mhchem}
\usepackage{pgfplots}
\usepackage[english]{babel}
\usepackage{amsmath}
\usepackage{amssymb}
\usepackage{stmaryrd}
\usepackage{xfrac}
\renewcommand{\tabref}{\Tabref}
\renewcommand{\figref}{\Figref}

\pgfplotsset{compat=1.3}
\usetikzlibrary{patterns}
\usetikzlibrary{plotmarks}

\@ifundefined{showcaptionsetup}{}{%
 \PassOptionsToPackage{caption=false}{subfig}}
\usepackage{subfig}
\makeatother

\begin{document}
\title{Transition-Potential Coupled Cluster}
\author{Megan Simons}
\author{Devin A. Matthews}
\email{damatthews@smu.edu}

\affiliation{Department of Chemistry, Southern Methodist University, Dallas, TX
75275}
\begin{abstract}
The problem of orbital relaxation in computational core-hole spectroscopies,
including x-ray absorption and x-ray photoionization, has long plagued
linear response approaches, including equation-of-motion coupled cluster
with singles and doubles (EOM-CCSD). Instead of addressing this problem
by including additional electron correlation, we propose an explicit
treatment of orbital relaxation via the use of ``transition potential''
reference orbitals, leading to a transition-potential coupled cluster
(TP-CC) family of methods. One member of this family in particular,
TP-CCSD($\sfrac{1}{2}$), is found to essentially eliminate the orbital
relaxation error and achieve the same level of accuracy for core-hole
spectra as is typically expected of EOM-CCSD in the valence region.
These results show that very accurate x-ray absorption spectra for
molecules with first-row atoms can be computed at a cost essentially
the same as that for EOM-CCSD.
\end{abstract}
\maketitle

\section{Introduction}

The use of x-ray spectroscopies has long been a mainstay in the study
of the structure and composition of ordered materials.\citep{heideXrayPhotoelectronSpectroscopy2011}
More recently, x-ray ionization, absorption, and (inelastic) scattering
have been applied to molecular systems, both in solution and in the
gas phase.\citep{koziejApplicationModernXray2017,normanSimulatingXraySpectroscopies2018}
Coupled cluster linear response (LR-CC,\citep{monkhorstCalculationPropertiesCoupledcluster1977,mukherjeeResponsefunctionApproachDirect1979,kochCoupledClusterResponse1990}
or equivalently equation-of-motion coupled cluster---EOM-CC\citep{sekinoLinearResponseCoupledcluster1984,stantonEquationMotionCoupledcluster1993,comeauEquationofmotionCoupledclusterMethod1993})
techniques are a natural method to apply to such ``core-ionized''
and ``core-excited'' (collectively ``core-hole'') states, given
their immense success in the description of valence excitations. Linear
response based on density functional theory (TD-DFT)\citep{grossTimeDependentDensityFunctionalTheory1990}
has been shown to describe core-hole states rather poorly because
of large self-interaction errors.\citep{brancatoAccurateDensityFunctional2008,normanSimulatingXraySpectroscopies2018}

Several density functionals have been tuned to reproduce x-ray absorption
spectra at the expense of a quantitative description of the ground
state electronic structure,\citep{tuSelfinteractioncorrectedTimedependentDensityfunctionaltheory2007,lestrangeCalibrationEnergySpecificTDDFT2015,jinAccurateComputationXray2018}
while range-separated functionals also somewhat decrease the self-interaction
error.\citep{besleyTimedependentDensityFunctional2009} Alternatively,
the $\Delta$Kohn-Sham method, which separately performs (quasi-)variational
optimizations of both the ground and core-hole states has been shown
to accurately reproduce absorption and ionization spectra in many
cases, but suffers from convergence difficulties and the need to compute
transition properties in a non-orthogonal framework.\citep{stenerDensityFunctionalCalculations1995,takahashiFunctionalDependenceCoreexcitation2004,besleySelfconsistentfieldCalculationsCore2009,evangelistaOrthogonalityConstrainedDensity2013,vermaPredictingEdgeXray2016,haitHighlyAccuratePrediction2020}
As a compromise, the family of ``transition-potential'' DFT (TP-DFT)
methods\citep{huDensityFunctionalComputations1996,trigueroCalculationsNearedgeXrayabsorption1998,trigueroSeparateStateVs1999,michelitschEfficientSimulationNearedge2019}
eschews both linear-response and state-specific orbital optimization
by performing a single calculation with fractional orbital occupation,
and is strongly motivated theoretically as an approximation to Slater's
Transition State method.

While TP-DFT has been successful in many cases, errors in both peak
positions and intensities remain that can prevent a firm assignment
of the spectrum in some cases. EOM-CC has been shown to smoothly and
rapidly reduce errors as the excitation level and basis set quality
are increased; in particular, inclusion of triple excitations and
the use of an augmented triple-zeta basis set (possibly with the addition
of explicit Rydberg basis functions) is sufficient for accuracy of
better than 0.1 eV (when relativistic effects are included) in absolute
energies.\citep{liuBenchmarkCalculationsKEdge2019,734c7d7e72234a548c61e39e075d7bd2}
The inclusion of triple excitations is particularly crucial as the
orbital relaxation effects for core-hole states are exceptionally
large. We have recently shown that a perturbative treatment of triple
excitations in the excited state (EOM-CCSD{*}) is sufficient to reproduce
the effect of full triple excitations to high accuracy.\citep{matthewsEOMCCMethodsApproximate2020}
Even so, the additional expense of triple excitations, especially
considering that such an effect is not typically required in valence
calculations at a similar level of accuracy, motivates the search
for an effective EOM-CC method that can accurately treat core-hole
states at a purely singles and doubles level.

In this paper, we present a potential candidate method, transition-potential
coupled cluster (TP-CC), that blends the best features of both TP-DFT
and EOM-CC in order to fulfill the need for an economical and yet
accurate EOM-CC method for treating core-hole states.

\section{Theoretical Methods}

\subsection{Equation-of-Motion Coupled Cluster Theory}

Equation-of-motion coupled cluster,\citep{sekinoLinearResponseCoupledcluster1984,stantonEquationMotionCoupledcluster1993,comeauEquationofmotionCoupledclusterMethod1993}
as well as the closely-related linear response coupled cluster theory,\citep{monkhorstCalculationPropertiesCoupledcluster1977,mukherjeeResponsefunctionApproachDirect1979,kochCoupledClusterResponse1990}
start with an exponential parametrization of the ground state,
\begin{align*}
E_{CC} & =\langle0|e^{-\hat{T}}\hat{H}_{N}e^{\hat{T}}|0\rangle\\
 & =\langle0|\bar{H}|0\rangle\\
0 & =\langle P|\bar{H}|0\rangle
\end{align*}
where the cluster operator $\hat{T}$ is a pure excitation operator
and $|0\rangle$ and $|P\rangle$ denote the reference determinant
and set of excited determinants, respectively. Practical coupled cluster
calculations require a truncation of the cluster operator and the
set of excited determinants; in this work, a truncation at the level
of single and double excitations (CCSD) is adopted. This leads to
the following definitions of $\hat{T}$ and $|P\rangle$, as well
as the standard definition of the normal-ordered Hamiltonian $\hat{H}_{N}$,
\begin{align*}
\hat{H}_{N} & =\sum_{pq}f_{q}^{p}\{p^{\dagger}q\}+\frac{1}{4}\sum_{pqrs}v_{rs}^{pq}\{p^{\dagger}q^{\dagger}sr\}\\
\hat{T} & =\sum_{ai}t_{i}^{a}a^{\dagger}i+\frac{1}{4}\sum_{abij}t_{ij}^{ab}a^{\dagger}b^{\dagger}ji\\
|P\rangle & =|S\rangle\oplus|D\rangle\\
 & =(a^{\dagger}i\,|0\rangle)\oplus(a^{\dagger}b^{\dagger}ji\,|0\rangle)
\end{align*}
We use the standard notations: $pqrs$ refer to creation/annihilation
operators of arbitrary spin-orbitals, while \textbf{$ab$} refer specifically
to virtual spin-orbitals and $ij$ to occupied spin-orbitals (with
respect to $|0\rangle$). Braces denote normal ordering; note that
$\hat{T}$ is implicitly normal ordered.

From this ground state, the $m$-th excited state is parametrized
by a linear excitation operator $\hat{R}_{m}$, a right eigenvector
of the similarity-transformed Hamiltonian $\bar{H}$, and the corresponding
excitation energy $\omega_{m}$ is determined as the corresponding
eigenvalue,
\begin{align*}
\omega_{m}\hat{R}_{m}|0\rangle & =(\bar{H}-E_{CC})\hat{R}_{m}|0\rangle\\
 & =[\bar{H},\hat{R}_{m}]|0\rangle\\
\hat{R}_{m} & =r_{0}+\sum_{ai}r_{i}^{a}a^{\dagger}i+\frac{1}{4}\sum_{abij}r_{ij}^{ab}a^{\dagger}b^{\dagger}ji
\end{align*}
The excitation energies obtained with EOM-CC are precisely equal to
the poles of the linear response function of the coupled cluster ground
state (LR-CC). These two theories only disagree in the definition
of transition properties such as oscillator strengths.\citep{kochCoupledClusterResponse1990}
In the length gauge, the EOM-CC dipole oscillator strength is given
as an expectation value of the dipole operator,
\begin{align*}
f_{m}(\text{EOM-CC}) & =\frac{2m_{e}\omega_{m}}{3\hbar^{2}}\sum_{\alpha=x,y,z}M_{m,\alpha}\\
M_{m,\alpha} & =\langle0|\hat{L}_{0}\bar{\mu}_{\alpha}\hat{R}_{m}|0\rangle\langle0|\hat{L}_{m}\bar{\mu}_{\alpha}\hat{R}_{0}|0\rangle
\end{align*}
where $\bar{\mu}_{\alpha}=e^{-\hat{T}}\hat{\mu}_{\alpha}e^{\hat{T}}$
and $\hat{\mu}_{\alpha}$ is the electronic dipole moment operator
along the $\alpha$ Cartesian axis. Because $\bar{H}$ is non-Hermitian,
it has distinct left eigenvectors $\hat{L}_{m}$. The ground state
eigenvectors are $\hat{R}_{0}=1$ and $\hat{L}_{0}=(1+\hat{\Lambda})$
with $\hat{\Lambda}$ being the usual coupled cluster amplitude response
operator. In the linear response formalism, the square transition
moments $M_{m,\alpha}$ are computed from the residual of the corresponding
pole in the response function. This gives rise to an additional term
that incorporates the response of the excitation amplitudes $r$ to
the electric field. In most circumstances this additional contribution
is very small and can be safely neglected, and we do so in this work.

In contrast to the DFT linear response (TD-DFT) formalism,\citep{grossTimeDependentDensityFunctionalTheory1990}
correlation of the excited state, which includes orbital relaxation
effects, is explicitly included in the excitation operator $\hat{R}_{m}$,
while in TD-DFT the exchange-correlation functional must account for
such effects. For core-hole states, the stark difference in length
scales between valence and core-hole correlation effects leads to
large self-interaction errors (SIEs) that fail to cancel between ground
and excited states. An explanation of these errors in terms of orbital
relaxation is also useful, as the ground state Kohn-Sham orbitals
that define the model system fail to sufficiently approximate the
core-hole state. While EOM-CC explicitly accounts for such effects,
an accurate treatment of them requires sufficient correlation in $\hat{R}_{m}$
which is only present at the triple excitation level.

When applied to core-excited and core-ionized states, which are in
fact resonances embedded in the valence ionization continuum, EOM-CC
typically encounters convergence problems. One solution to this problem
is the core-valence separation (CVS), first introduced by Cederbaum
and Schirmer,\citep{cederbaumManybodyTheoryCore1980,barthTheoreticalCorelevelExcitation1985}
and adapted to EOM-CC by Coriani and Koch.\citep{corianiCommunicationXrayAbsorption2015}
In this approach, pure valence excitations or ionizations are excluded
from the linear response manifold, which both restores convergence
and eliminates spurious couplings to the (badly) discretized continuum
determinants.\citep{liuBenchmarkCalculationsKEdge2019} All EOM-CC
methods considered here use the CVS.

\subsection{Transition-Potential Density Functional Theory}

The TP-DFT theory\citep{huDensityFunctionalComputations1996,trigueroCalculationsNearedgeXrayabsorption1998,trigueroSeparateStateVs1999,michelitschEfficientSimulationNearedge2019}
is an approximation to Slater's Transition State (TS) method, which
in turn is ultimately derived from $\Delta$Kohn-Sham (or $\Delta$DFT).
In the latter approach, separate DFT calculations are performed for
the initial and final states, and the resulting energies are simply
subtracted,
\begin{align*}
\omega_{\Delta KS} & =E_{f}-E_{i}
\end{align*}
Given a suitable homotopy that connects the orbitals of the initial
and final states by a continuous parameter $\lambda$ (with $\lambda=0$
in the initial state and $\lambda=1$ in the final state), the energy
difference can be written as,
\begin{align*}
\omega_{\Delta KS} & =\int_{0}^{1}\frac{dE(\lambda)}{d\lambda}d\lambda
\end{align*}
Now, assume that the initial and final states differ only by a single
excitation, that is, we can identify a (spin-)orbital $\phi_{1}$
that has an occupation $n_{1}=1$ in the initial state and 0 in the
final state, and another orbital $\phi_{2}$ that has occupation $n_{2}=0$
in the initial state and 1 in the final state, while all other orbitals
have the same occupation (1 or 0) in both states. Note that the actual
(spatial) orbitals need not be the same in both states, only that
they can be uniquely identified via the homotopy. Thus,
\begin{align*}
\frac{dE(\lambda)}{d\lambda} & =\frac{\partial E(\lambda)}{\partial n_{1}}\frac{\partial n_{1}}{\partial\lambda}+\frac{\partial E(\lambda)}{\partial n_{2}}\frac{\partial n_{2}}{\partial\lambda}\\
 & =-\frac{\partial E(\lambda)}{\partial n_{1}}+\frac{\partial E(\lambda)}{\partial n_{2}}
\end{align*}
Janak's theorem then provides the necessary partial derivatives $\partial E/\partial n_{i}=\epsilon_{i}$
from which we can arrive at,
\begin{align*}
\omega_{\Delta KS} & =\int_{0}^{1}\left[\epsilon_{2}(\lambda)-\epsilon_{1}(\lambda)\right]d\lambda\\
 & \approx\epsilon_{2}(1/2)-\epsilon_{1}(1/2)
\end{align*}
where the second step is Slater's TS which is the first-order approximation
to the exact energy difference (the mid-point rule).

The TS method typically provides a good estimate of the excitation
energies, even for core-hole states, but is complicated by the need
to converge a half-electron state (with $n_{1}=n_{2}=1/2$), especially
with regard to the partial occupation of the virtual orbital. In a
sufficiently diffuse basis set, typical methods to converge such a
state, e.g. using the maximum overlap method, are prone to failure.
For high-quality $\Delta$KS calculations, more elaborate methods
such as orthogonality-constrained DFT or constrained variational excited
state optimization can be employed. Instead, the transition-potential
(TP) approach further approximates TS by setting the virtual orbital
occupation to zero, i.e. $\omega_{\Delta KS}\approx\epsilon_{2}(n_{1}=1/2,n_{2}=0)-\epsilon_{1}(n_{1}=1/2,n_{2}=0)$.
The advantages of TP over TS are 1) the half-electron state (in this
case a half-core-hole---HCH---state) can be more reliably converged,
and 2) the same half-electron state may be used for excitations to
any virtual orbital. In the context of XAS (NEXAFS), this means that
the entire spectrum due to excitation of a particular core orbital
may be obtained in a single calculation.

The selection of a half-electron state is derived via simple one-point
approximation of the energy difference integral, but it can also be
rationalized from an error cancellation perspective. In TD-DFT, the
ground state can be considered well-described (i.e. described as well
as the chosen exchange-correlation functional is capable of), but
the description of the excited state is hampered by orbital relaxation/SIE.
If we were to start from the optimized core-hole state and compute
the ground state energy using linear response, the opposite would
be true. However, by starting from the half-core-hole state, errors
in both directions are largely canceled.

While the TS/TP-DFT method computes the energies using well-justified
approximations to the exact $\Delta KS$ energy difference, there
is not a corresponding set of approximations for the transition moments.
Typically, a simple formula based on the sudden approximation is employed,
\begin{align*}
f_{1\shortrightarrow2}(\text{TP-DFT}) & =\frac{2m_{e}(\epsilon_{2}-\epsilon_{1})}{3\hbar^{2}}\sum_{\alpha=x,y,z}M_{1\shortrightarrow2,\alpha}\\
M_{1\shortrightarrow2,\alpha} & =2\left|\langle\phi_{2}|\hat{\mu}_{\alpha}|\phi_{1}\rangle\right|^{2}
\end{align*}
Instead of a specific state label $m$, the particular transition
is determined by the orbitals $\phi_{1}$ and $\phi_{2}$. Because
only the core spin-orbital of one spin (typically $\beta$ spin) is
half-occupied, the factor of two in $M_{1\shortrightarrow2,\alpha}$
is necessary to account for the ``missing'' $\alpha$ spin component
which is identical due to spin symmetry for a closed-shell reference
configuration. Starting from an unrestricted open-shell configuration,
distinct $\alpha$ and $\beta$ excitation spectra would need to be
computed.

\subsection{Transition-Potential Coupled Cluster}

The error cancellation perspective on the TP-DFT approach suggests
a possible route for ameliorating the orbital relaxation error in
EOM-CC as well. CVS-EOM-CCSD reliably overshoots both vertical core-excitation
and core-ionization energies by 1--2 eV (see results below). Thus,
the effect of triple excitations, which largely eliminates the orbital
relaxation error, always acts to stabilize the final state. Now, let
us examine the effect of substituting the ground state Hartree--Fock
orbitals with another set of orbitals that explicitly include some
amount of core-hole relaxation: first, the use of non-Hartree--Fock
orbitals of course raises the reference energy, as the HF orbitals
are variationally optimized; while CC is not variational, arbitrary
changes to the orbitals also typically raise the CC energy. This should
be especially true for highly non-optimal orbitals that include core
relaxation. Second, the explicit inclusion of core relaxation should
stabilize the final core-excited or core-ionized state at the CCSD
level, in a similar manner as the inclusion of triple excitations.
This stabilization effect is expected to increase in proportion to
the amount of explicit relaxation included in the orbitals.

The first effect is, in isolation, a degradation in the physical description
of the ground state. However, since the increase in energy of the
ground state has the same effect on the vertical energy differences
as a lowering of the final state, both effects of substituting the
orbitals in practice act in concert. Therefore, there should be some
set of \emph{partially-relaxed} orbitals that combines \emph{partial
destabilization} of the ground state with \emph{partial stabilization}
of the excited state that, combined, reproduce the full orbital relaxation
effect, but at the CCSD level. To this end, we have defined a family
of ``transition-potential coupled cluster'' (TP-CC) methods, which
vary in two ways. First, the choice of how much relaxation to include
in the orbitals is parametrized by $\lambda$ as in the previous section.
Second, the particular orbitals are obtained from a fractional-occupation
SCF (here B3LYP) calculation with either a partial core hole (as in
TP-DFT), or a partial core excitation to a virtual orbital (which
we call XTP as in Ref.~\citenum{michelitschEfficientSimulationNearedge2019}).
These TP-CCSD($\lambda$) and XTP-CCSD($\lambda$) methods, apart
from the non-standard choice of orbitals, are simply standard CVS-EOM-CCSD
calculations.

\section{Computational Details}

The (X)TP-CCSD($\lambda$) methods were implemented via a combination
of the Psi4\citep{smithPSI4OpensourceSoftware2020} and CFOUR\citep{matthewsCoupledclusterTechniquesComputational2020}
program packages. Specifically, we utilized the PSIXAS plugin\citep{ehlertPSIXASPsi4Plugin2020a}
for Psi4 to generate fractional core-hole or core-excited orbitals,
using the B3LYP functional and ionizing or exciting $\lambda/2$ electrons
of each spin in the selected core orbital. Excited electrons were
promoted to the LUMO in each case. We modified the PSIXAS plugin to
produce basis set ($\texttt{GENBAS}$) and molecular orbital ($\texttt{OLDMOS}$)
files suitable for use in CFOUR. We also modified the CFOUR symmetry
analysis code to prevent any reorientation or translation of the molecule
in order to exactly match Psi4's reference frame. In CFOUR, the reference
orbitals were first reoccupied in the standard Aufbau ordering. Since
the resulting determinant is clearly not a Hartree--Fock solution
we included the $\texttt{NON-HF=ON}$ keyword in the CFOUR input file,
and also requested semicanonicalization of the orbitals. Then, a standard
CVS-EOM-CCSD calculation is performed, including all non-Hartree--Fock
terms. Oscillator strengths were calculated using the expectation
value formalism described above.

The test set consisted of all 1s principal core ionizations and four
core excitations (for each 1s orbital) of $\ce{H2O}$, CO, HCN, HF,
HOF, HNO, $\ce{CH2}$, $\ce{CH4}$, $\ce{NH3}$, $\ce{H3CF}$ $\ce{H3COH}$,
$\ce{H2CO}$, $\ce{H2CNH}$, and $\ce{H2NF}$. The core excitations
were selected as those for which we could reliably converge all methods
tested, which typically consisted of the first four excitations of
dominant single excitation character. Double excitations were specifically
avoided as EOM-CCSD is known to describe them quite poorly even in
the valence case. All calculations utilized the aug-cc-pCVTZ basis
set with all electrons correlated, except for $\ce{H2O}$ where aug-cc-pCVQZ
was used. In order to avoid complications due to missing relativistic
effects, basis set incompleteness (particularly for Rydberg core excitations),
geometric effects, and data quality and availability, which would
all be a concern when comparing directly to experimental data, we
have used full CVS-EOM-CCSDT as a benchmark. Carbone et al.\citealp{734c7d7e72234a548c61e39e075d7bd2}
showed that CCSDT is typically within 100 meV of the experimental
(vertical) core excitation energies, while Liu. et al.\citep{liuBenchmarkCalculationsKEdge2019}
showed similar results for core ionization potentials. In addition
to CVS-EOM-CCSD, TP-CCSD($\sfrac{1}{2}$), TP-CCSD($\sfrac{1}{4}$),
XTP-CCSD($\sfrac{1}{2}$), and XTP-CCSD($\sfrac{1}{4}$) values, we
have also expanded the CVS-EOM-CCSD{*} results from Ref.~\citenum{matthewsEOMCCMethodsApproximate2020}
to encompass the larger test set used here. These results are included
as an ``aspirational yardstick'', since that method was previously
found to reproduce the full CVS-EOM-CCSDT results rather well.

\section{Results and Discussion}

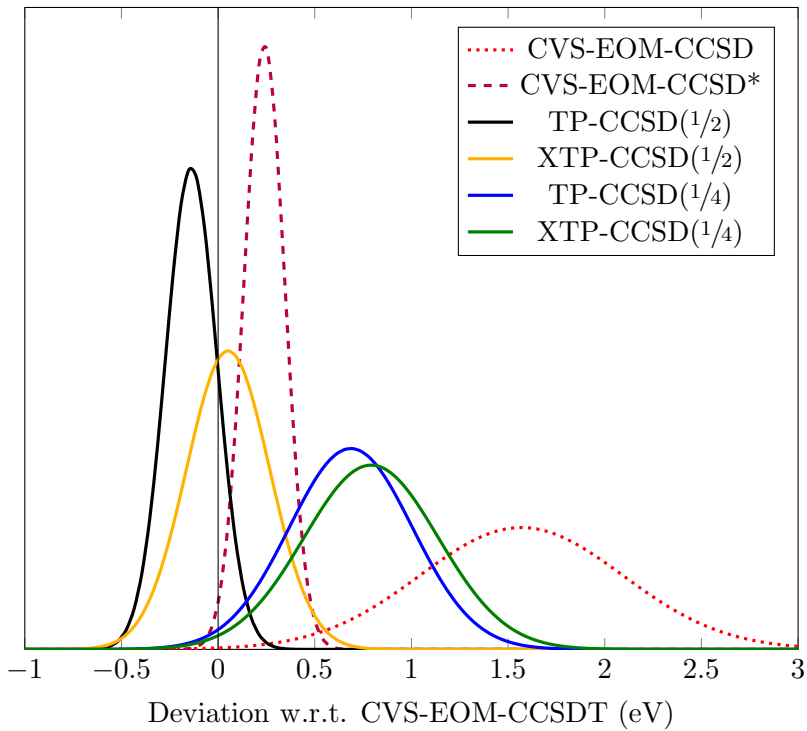
\begin{figure}
\begin{tikzpicture}
\begin{axis}[
scale=1.5,smooth,
xmin=-1, xmax=3,
ymin=0, ymax=4,
ytick=\empty,
xlabel={Deviation w.r.t. CVS-EOM-CCSDT (eV)},
legend style={legend pos=north east}
]
\addplot[draw=red,very thick,dotted] table[x=x,y=ccsd] {abs-energy.dat};
\addlegendentry{CVS-EOM-CCSD}
\addplot[draw=purple,very thick,dashed] table[x=x,y=ccsd*] {abs-energy.dat};
\addlegendentry{CVS-EOM-CCSD*}
\addplot[draw=black,very thick] table[x=x,y=tphalf] {abs-energy.dat};
\addlegendentry{TP-CCSD($\sfrac{1}{2}$)}
\addplot[draw=orange!60!yellow,very thick] table[x=x,y=xtphalf] {abs-energy.dat};
\addlegendentry{XTP-CCSD($\sfrac{1}{2}$)}
\addplot[draw=blue,very thick] table[x=x,y=tpquarter] {abs-energy.dat};
\addlegendentry{TP-CCSD($\sfrac{1}{4}$)}
\addplot[draw=black!50!green,very thick] table[x=x,y=xtpquarter] {abs-energy.dat};
\addlegendentry{XTP-CCSD($\sfrac{1}{4}$)}
\addplot[draw=black] coordinates {(0,-1) (0,100)};
\end{axis}
\end{tikzpicture}

\caption{\label{fig:absolute}Normal error distributions for core excitation
energies. Iterative EOM-CC methods are denoted by dotted lines, perturbative
EOM-CC methods by dashed lines, and TP-CC methods by solid lines.}
\end{figure}
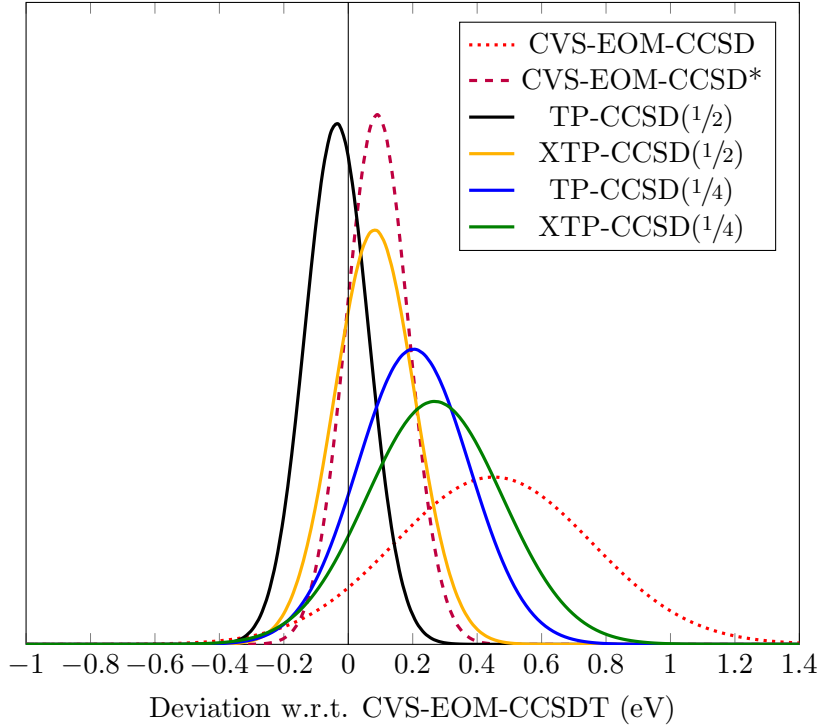
\begin{figure}
\begin{tikzpicture}
\begin{axis}[
scale=1.5,smooth,
xmin=-1, xmax=1.4,
ymin=0, ymax=5,
ytick=\empty,
xlabel={Deviation w.r.t. CVS-EOM-CCSDT (eV)},
legend style={legend pos=north east}
]
\addplot[draw=red,very thick,dotted] table[x=x,y=ccsd] {rel-energy.dat};
\addlegendentry{CVS-EOM-CCSD}
\addplot[draw=purple,very thick,dashed] table[x=x,y=ccsd*] {rel-energy.dat};
\addlegendentry{CVS-EOM-CCSD*}
\addplot[draw=black,very thick] table[x=x,y=tphalf] {rel-energy.dat};
\addlegendentry{TP-CCSD($\sfrac{1}{2}$)}
\addplot[draw=orange!60!yellow,very thick] table[x=x,y=xtphalf] {rel-energy.dat};
\addlegendentry{XTP-CCSD($\sfrac{1}{2}$)}
\addplot[draw=blue,very thick] table[x=x,y=tpquarter] {rel-energy.dat};
\addlegendentry{TP-CCSD($\sfrac{1}{4}$)}
\addplot[draw=black!50!green,very thick] table[x=x,y=xtpquarter] {rel-energy.dat};
\addlegendentry{XTP-CCSD($\sfrac{1}{4}$)}
\addplot[draw=black] coordinates {(0,-1) (0,100)};
\end{axis}
\end{tikzpicture}

\caption{\label{fig:rel}Normal error distributions for core excitation relative
values. Iterative EOM-CC methods are denoted by dotted lines, perturbative
EOM-CC methods by dashed lines, and TP-CC methods by solid lines.}
\end{figure}
\begin{figure}
\begin{tikzpicture}
\begin{axis}[
scale=1.5,smooth,
xmin=-1, xmax=3,
ymin=0, ymax=5,
ytick=\empty,
xlabel={Deviation w.r.t. CVS-EOM-CCSDT (eV)},
legend style={legend pos=north east}
]
\addplot[draw=red,very thick,dotted] table[x=x,y=ccsd] {ionization.dat};
\addlegendentry{CVS-EOM-CCSD}
\addplot[draw=purple,very thick,dashed] table[x=x,y=ccsd*] {ionization.dat};
\addlegendentry{CVS-EOM-CCSD*}
\addplot[draw=black,very thick] table[x=x,y=tphalf] {ionization.dat};
\addlegendentry{TP-CCSD($\sfrac{1}{2}$)}
\addplot[draw=orange!60!yellow,very thick] table[x=x,y=xtphalf] {ionization.dat};
\addlegendentry{XTP-CCSD($\sfrac{1}{2}$)}
\addplot[draw=blue,very thick] table[x=x,y=tpquarter] {ionization.dat};
\addlegendentry{TP-CCSD($\sfrac{1}{4}$)}
\addplot[draw=black!50!green,very thick] table[x=x,y=xtpquarter] {ionization.dat};
\addlegendentry{XTP-CCSD($\sfrac{1}{4}$)}
\addplot[draw=black] coordinates {(0,-1) (0,100)};
\end{axis}
\end{tikzpicture}

\caption{\label{fig:ip}Normal error distributions for core ionization energies.
Iterative EOM-CC methods are denoted by dotted lines, perturbative
EOM-CC methods by dashed lines, and TP-CC methods by solid lines.}
\end{figure}
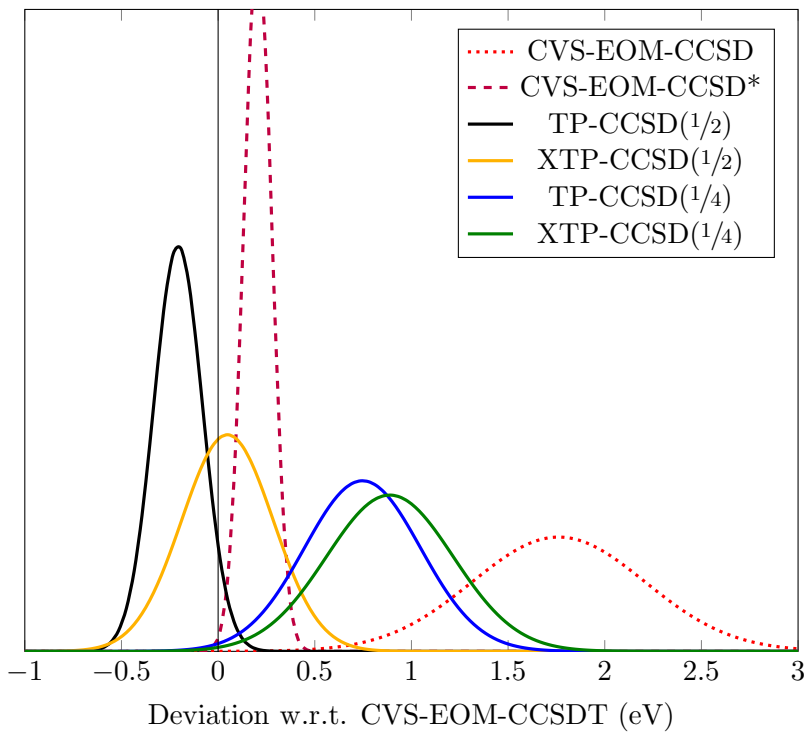

In the following discussion, the ``shortened'' names of the CVS-EOM
methods will be used, e.g. CCSDT = CVS-EOM-CCSDT. The distribution
of ``absolute'' (i.e. unmodified vertical) excitation energy deviations
from CCSDT are depicted in \figref{absolute}. The absolute energy
deviation for a method $X$ is calculated as $E(X)-E(\text{CCSDT})$
where $E$ is a vertical core excitation energy or core ionization
potential. The ``relative'' excitation energy deviations are depicted
in \figref{rel}. These deviations are determined from excitation
energies adjusted such that the lowest excitation energy for each
edge is equal to 0 (this is essentially a shift of the entire spectrum;
note that the shift is different for each method, and is applied before
computing the deviations). One relative core excitation out of four
is trivially zero after adjustment; these values are not included
in the statistics. Since the methods should make similar errors in
the ionization potential energies and excitation energies, the relative
errors should be smaller due to error cancellation. A similar shift
is commonly applied when comparing to experimental data. Finally,
the distribution of computed ionization potentials is depicted in
\figref{ip}. In each figure, the distribution of the energy deviations
is fit to a Gaussian (normal distribution).

\subsection{CCSD and CCSD{*}}

As reported in previous studies,\citep{corianiCommunicationXrayAbsorption2015,734c7d7e72234a548c61e39e075d7bd2,vidalNewEfficientEquationofMotion2019,matthewsEOMCCMethodsApproximate2020}
CCSD systematically overestimates all core excitation energies. The
large orbital relaxation energy is challenging for a purely linear
response method due to the localized nature of the core hole, and
absolute errors of 1--3 eV remain at the CCSD level. In comparison,
EOM-CCSD typically reproduces EOM-CC3 vertical valence excitation
energies to within 0.3 eV.\citep{schreiberBenchmarksElectronicallyExcited2008}
A simple triples correction to the excited (core-hole) state only
(CCSD{*}) nearly eliminates the deviation with respect to CCSDT, indicating
that the leading high-order correlation effects in the upper state
(which correspond to orbital relaxation) are the primary source of
error. Residual errors in CCSD{*} are potentially indicative of improvement
in the correlation of both the ground and excited states, and are
of a similar magnitude to triples effects in valence excitations,
and so the CCSD{*} values likely represent a ``best estimate'' of
the effect of orbital relaxation alone.

\subsection{TP-CCSD($\sfrac{1}{2}$) and XTP-CCSD($\sfrac{1}{2}$)}

The choice of $\ensuremath{\lambda=\sfrac{1}{2}}$ can be expected
to be a reasonable first-order estimate of the optimal error-cancellation
point for TP-CC methods. Note that while a half-core-hole is provably
optimal as a single-point approximation in the TS- and TP-DFT methods,
there is no such formal argument for TP-CC. Nonetheless, these calculations
show significantly better agreement with CCSDT in comparison to the
CCSD calculations with standard Hartree-Fock orbitals. The absolute
error distributions can be seen in \figref{absolute}, where the distribution
for TP-CCSD($\sfrac{1}{2}$) has a similar shape to CCSD{*} but with
slightly smaller average error. XTP-CCSD($\sfrac{1}{2}$) reduces
the average error even more, to below 0.1 eV, but exhibits approximately
twice the variability of CCSD{*}. \figref{rel} shows the relative
error distribution, where the distribution for XTP-CCSD($\sfrac{1}{2}$)
is almost identical to CCSD{*} but again with slightly lower average
error. XTP-CCSD($\sfrac{1}{2}$) provides similar statistical deviations,
although the standard deviation is slightly increased over both TP-CCSD($\sfrac{1}{2}$)
and CCSD{*}. Similarly to the absolute excitation energies, both TP-CCSD($\sfrac{1}{2}$)
and XTP-CCSD($\sfrac{1}{2}$) show an improvement over CCSD for core
ionization potentials, with XTP-CCSD($\sfrac{1}{2}$) again showing
a smaller average error but larger standard deviation in comparison
with TP-CCSD($\sfrac{1}{2}$). The ionization potential distribution
for these methods is shown in \figref{ip}. CCSD{*} and TP-CCSD($\sfrac{1}{2}$)
have very similar distributions, but TP-CCSD($\sfrac{1}{2}$) does
not attain quite as small a standard deviation. As with excitation
energies, evaluating relative ionization energies, specifically ionization
``chemical shifts''\citep{liuBenchmarkCalculationsKEdge2019} relative
to a standard, may further reduce the average and standard deviation
of the error.

Both TP-CC methods significantly improve on CCSD, and for shifted
spectra, account for essentially all of the orbital relaxation energy
(using CCSD{*} as a yardstick). Additionally, the improvement when
considering shifted rather than absolute spectra shows that TP-CC
additionally benefits from error cancellation within the spectrum.
While CCSD also displays the same benefit, there are still residual
errors as large as 1 eV, while TP-CC maintains deviation from full
CCSDT within 0.3 eV, closely matching the performance of CCSD in the
valence region. The XTP-CC variant was introduced with the idea that
the use of a neutral reference state for determining the orbitals
should balance a tendency toward over-contraction that might be expected
from a fractionally-charged system. However, at $\ensuremath{\lambda=\sfrac{1}{2}}$
this seems not to be the case. Upon a closer investigation of the
individual data (see Supporting Information), it can be seen that
XTP-CC does in fact slightly improve the description of valence resonances
(e.g. $\text{1s}\rightarrow\pi^{*}$), but slightly worsens the description
of Rydberg states (the standard deviation is increased by $\sim80\%$,
although the mean error is actually reduced). Rydberg states are less
sensitive to correlation and a balanced description of valence and
Rydberg states is a hallmark of a ``good'' method. XTP-CC, where
the fractional electron is placed in the LUMO, which is typically
a valence anti-bonding orbital, seems to lose this balance.

The simple choice of $\ensuremath{\lambda=\sfrac{1}{2}}$ seems to
do a remarkably good job of almost entirely eliminating the orbital
relaxation error for both core excitation energies and core ionization
potentials. However, a finer tuning of the $\lambda$ parameter may
further decrease either the average error or variability (standard
deviation) of the method. In particular, TP-CCSD($\sfrac{1}{2}$)
seems to slightly underestimate absolute energies while CCSD overestimates,
indicating that a slightly smaller value of $\lambda$ might offer
further improvement. While it would be impractical to tune $\lambda$
for each individual system (and largely defeat the purpose of an \emph{ab
initio} method like EOM-CC), a close inspection of the results shows
that, for example, the standard deviation of the core excitation energies
does increase monotonically on going from C to F. While this, to some
extent, reflects the increase in energy scale, it may also indicate
that different elements require sightly different optimal $\lambda$
values. We will explore optimization of $\lambda$ in a future publication,
but for now it seems that $\ensuremath{\lambda=\sfrac{1}{2}}$ is
a good default value.

\subsection{TP-CCSD($\sfrac{1}{4}$) and XTP-CCSD($\sfrac{1}{4}$)}

In addition to $\ensuremath{\lambda=\sfrac{1}{2}}$, we also tried
$\ensuremath{\lambda=\sfrac{1}{4}}$, $\ensuremath{\lambda=\sfrac{3}{4}}$,
and even $\ensuremath{\lambda=1}$ TP-CC calculations, in order to
understand the tradeoff between ground state destabilization and final
state stabilization. The latter two choices resulted in a lack of
convergence of the ground state coupled cluster equations in all cases.
This is not surprising, given that the choice of very different reference
orbitals will induce large cluster amplitudes. In the case of $\ensuremath{\lambda=1}$,
we would even expect $\hat{T}_{1}$ amplitudes on the order of 1,
which would completely destabilize the (truncated) coupled cluster
procedure. While $\ensuremath{\lambda=\sfrac{1}{2}}$ is clearly a
good choice for TP-CC, it is worthwhile to also examine $\ensuremath{\lambda=\sfrac{1}{4}}$.

The absolute excitation energy distributions (\figref{absolute})
for TP-CCSD($\sfrac{1}{4}$) and XTP-CCSD($\sfrac{1}{4}$) sit nicely
between those of CCSD and CCSD{*}/(X)TP-CCSD($\sfrac{1}{2}$), both
in terms of average error as well as standard deviation. This shows
that there is still some error cancellation between the initial and
final state effects, although the cancellation is incomplete. The
relationship of XTP-CCSD($\sfrac{1}{4}$) to TP-CCSD($\sfrac{1}{4}$)
seems to mirror that of XTP-CCSD($\sfrac{1}{2}$) and TP-CCSD($\sfrac{1}{2}$),
with the former in each case having a slightly higher standard deviation.
Moving to relative excitation energies (\figref{rel}) shows a very
similar situation, as does looking at ionization energies (\figref{ip}).
In each case, XTP-CCSD($\sfrac{1}{4}$) performs slightly worse than
its TP-CCSD($\sfrac{1}{4}$) counterpart. In contrast to $\lambda=\sfrac{1}{2}$,
where XTP-CC did in fact decrease the average error for valence and
even Rydberg states (although it simultaneously increased the standard
deviation of the latter), the average errors for all states increase
by approximately 0.1 eV in XTP-CCSD($\sfrac{1}{4}$) compared to TP-CCSD($\sfrac{1}{4}$).
The standard deviations are also slightly higher across the board,
although only very slightly unlike with XTP-CCSD($\sfrac{1}{2}$).
Although we have only sampled $\ensuremath{\lambda=\sfrac{1}{4}}$
and $\ensuremath{\lambda=\sfrac{1}{2}}$ thus far, it seems that the
error cancellation properties of TP-CC vary in fairly direct proportion
with the fraction of electron ionized in the core orbital.

\subsection{Oscillator Strengths}

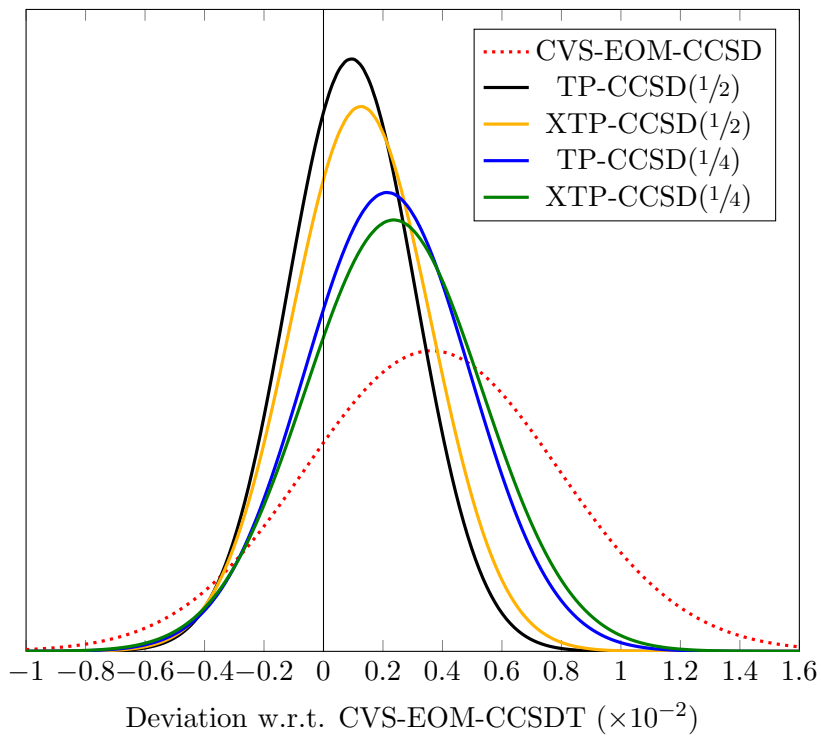
\begin{figure}
\begin{tikzpicture}

\begin{axis}[
scale=1.5,smooth,
xmin=-0.01, xmax=0.016,
ymin=0, ymax=200,
ytick=\empty,
xlabel={Deviation w.r.t. CVS-EOM-CCSDT ($\times 10^{-2}$)},
legend style={legend pos=north east},
xtick scale label code/.code={}
]
\addplot[draw=red,very thick,dotted] table[x=x,y=ccsd] {abs-int.dat};
\addlegendentry{CVS-EOM-CCSD}
\addplot[draw=black,very thick] table[x=x,y=tphalf] {abs-int.dat};
\addlegendentry{TP-CCSD($\sfrac{1}{2}$)}
\addplot[draw=orange!60!yellow,very thick] table[x=x,y=xtphalf] {abs-int.dat};
\addlegendentry{XTP-CCSD($\sfrac{1}{2}$)}
\addplot[draw=blue,very thick] table[x=x,y=tpquarter] {abs-int.dat};
\addlegendentry{TP-CCSD($\sfrac{1}{4}$)}
\addplot[draw=black!50!green,very thick] table[x=x,y=xtpquarter] {abs-int.dat};
\addlegendentry{XTP-CCSD($\sfrac{1}{4}$)}
\addplot[draw=black] coordinates {(0,-1) (0,200)};
\end{axis}
\end{tikzpicture}

\caption{\label{fig:absolute-int}Normal error distributions for absolute oscillator
strengths. Iterative EOM-CC methods are denoted by dotted lines, and
TP-CC methods by solid lines.}
\end{figure}
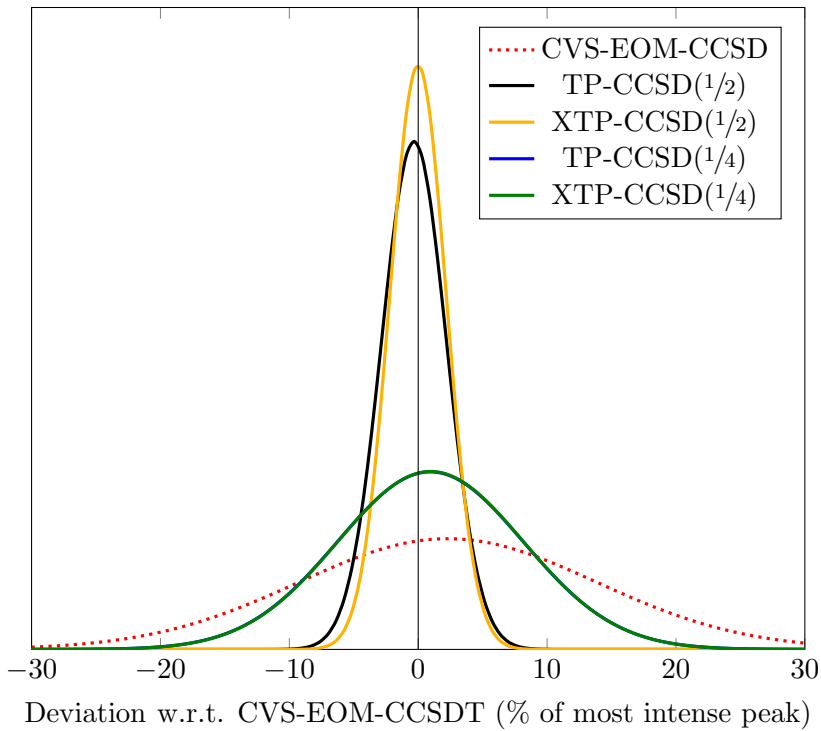
\begin{figure}
\begin{tikzpicture}
\begin{axis}[
scale=1.5,smooth,
xmin=-30, xmax=30,
ymin=0, ymax=0.2,
ytick=\empty,
xlabel={Deviation w.r.t. CVS-EOM-CCSDT (\% of most intense peak)},
legend style={legend pos=north east}
]
\addplot[draw=red,very thick,dotted] table[x=x,y=ccsd] {rel-int.dat};
\addlegendentry{CVS-EOM-CCSD}
\addplot[draw=black,very thick] table[x=x,y=tphalf] {rel-int.dat};
\addlegendentry{TP-CCSD($\sfrac{1}{2}$)}
\addplot[draw=orange!60!yellow,very thick] table[x=x,y=xtphalf] {rel-int.dat};
\addlegendentry{XTP-CCSD($\sfrac{1}{2}$)}
\addplot[draw=blue,very thick] table[x=x,y=tpquarter] {rel-int.dat};
\addlegendentry{TP-CCSD($\sfrac{1}{4}$)}
\addplot[draw=black!50!green,very thick] table[x=x,y=xtpquarter] {rel-int.dat};
\addlegendentry{XTP-CCSD($\sfrac{1}{4}$)}
\addplot[draw=black] coordinates {(0,-1) (0,1)};
\end{axis}
\end{tikzpicture}

\caption{\label{fig:rel-int}Normal error distributions for relative oscillator
strengths. Iterative EOM-CC methods are denoted by dotted lines, and
TP-CC methods by solid lines. The TP-CCSD($\sfrac{1}{4}$) and XTP-CCSD($\sfrac{1}{4}$)
curves overlap.}
\end{figure}

Because TP-CC is, computationally, identical to a standard EOM-CC
calculation, it is simple to compute oscillator strengths, here in
the expectation value formalism. We have also added full EOM-CCSDT
transition properties in the development version of CFOUR, and so
we can benchmark the effect of transition-potential orbitals on this
important property. EOM-CCSD{*}, as a perturbative correction to EOM-CCSD,
does not provide corrected oscillator strengths. The absolute deviations
in the dimensionless oscillator strengths, defined in analogy to the
absolute excitation and ionization energy deviations, are depicted
statistically in \figref{absolute-int}. In addition, we have computed
the deviations for relative oscillator strengths, which have been
normalized separately for each spectrum such that the most intense
transition has unit strength; the statistics for these relative deviations
are depicted in \figref{rel-int} as percentages.

The deviations of absolute oscillator strengths (\figref{absolute-int})
seem to indicate a similar improvement over CCSD as seen in the case
of absolute excitation energies. The average errors decrease sharply
on going from CCSD to (X)TP-CCSD($\sfrac{1}{4}$) to (X)TP-CCSD($\sfrac{1}{2}$),
with a fairly linear dependence on $\lambda$. Additionally, the XTP-CC
methods evince slightly higher standard deviations as in the case
of the energies. However, these statistics are largely dominated by
the more intense peaks in the spectrum, often the excitations to valence
anti-bonding orbitals. Instead, looking at the rescaled intensities
(\figref{rel-int}) shows a rather different effect. Here, the deviations
represent the normalized deviation, with the most intense peak set
at unity. This measure was chosen rather than relative intensity changes
for each peak, because in the latter case large relative changes of
very weak transitions (which are ultimately much less important to
the overall assignment) would dominate the statistics. Instead, a
spectrum-wide normalization, like the spectrum-wide energy shift in
\figref{rel}, puts all spectra on an equal footing while preserving
the relative importance of each peak in the spectrum. In this case,
we can see that the improvement of the TP-CC methods over CCSD is
even more pronounced, with a 3--4$\times$ reduction in standard
deviation, and essentially zero average deviation. Interestingly,
XTP-CCSD($\sfrac{1}{2}$) now performs very slightly better than TP-CCSD($\sfrac{1}{2}$).

The significant improvement of TP-CC oscillator strengths over CCSD
is very encouraging, as obtaining the correct relative intensities
is often nearly as important to fully assigning a spectrum as the
transition energies. While the calculations here do not include the
necessary diffuse orbitals in order to directly compare the Rydberg
states to experiment, it is most likely that the improved performance
of TP-CC will prove important for properly predicting Rydberg spectra,
in addition to the usually more intense valence peaks. The improvement
of TP-CC oscillator strengths also suggests that other one-electron
properties, such as multipole moments, may also be improved relative
to CCSD, although this is beyond the scope of the current study.

\subsection{Statistical Analysis}

\begin{table}
\subfloat[\label{tab:Normality-values-for}``Normality'' values for each data
category. See text for details.]{%
\begin{tabular}{|c|>{\centering}p{0.15\textwidth}|>{\centering}p{0.15\textwidth}|>{\centering}p{0.15\textwidth}|>{\centering}p{0.15\textwidth}|>{\centering}p{0.15\textwidth}|}
\cline{2-6} \cline{3-6} \cline{4-6} \cline{5-6} \cline{6-6} 
\multicolumn{1}{c|}{} & Absolute Excitation Energies

($n=92$) & Relative Excitation Energies

($n=69$) & Ionization Potentials

($n=23$) & Absolute Oscillator Strengths ($n=90$) & Relative Oscillator Strengths ($n=67$)\tabularnewline
\hline 
CVS-EOM-CCSD & 0.97 & 0.97 & 0.95 & \textbf{\emph{0.90}} & \textbf{\emph{0.59}}\tabularnewline
\hline 
CVS-EOM-CCSD{*} & 0.97 & 0.94 & 0.96 & -- & --\tabularnewline
\hline 
TP-CCSD($\sfrac{1}{2}$) & 0.99 & 0.98 & 0.97 & \textbf{\emph{0.73}} & \textbf{\emph{0.77}}\tabularnewline
\hline 
XTP-CCSD($\sfrac{1}{2}$) & 0.98 & \textbf{\emph{0.88}} & 0.95 & \textbf{\emph{0.72}} & \textbf{\emph{0.80}}\tabularnewline
\hline 
TP-CCSD($\sfrac{1}{4}$) & 0.95 & 0.94 & 0.93 & \textbf{\emph{0.87}} & \textbf{\emph{0.59}}\tabularnewline
\hline 
XTP-CCSD($\sfrac{1}{4}$) & 0.96 & 0.96 & 0.96 & \textbf{\emph{0.88}} & \textbf{\emph{0.60}}\tabularnewline
\hline 
\end{tabular}

}

\subfloat[Estimated skewness values for each distribution.]{%
\begin{tabular}{|c|>{\centering}p{0.15\textwidth}|>{\centering}p{0.15\textwidth}|>{\centering}p{0.15\textwidth}|>{\centering}p{0.15\textwidth}|>{\centering}p{0.15\textwidth}|}
\hline 
CVS-EOM-CCSD & 0.33 & 0.12 & 0.24 & \textbf{\emph{0.90}} & \textbf{\emph{3.39}}\tabularnewline
\hline 
CVS-EOM-CCSD{*} & 0.24 & \textbf{\emph{0.69}} & \textbf{\emph{-0.58}} & -- & --\tabularnewline
\hline 
TP-CCSD($\sfrac{1}{2}$) & 0.26 & 0.42 & \textbf{\emph{0.52}} & \textbf{\emph{1.94}} & \textbf{\emph{-0.97}}\tabularnewline
\hline 
XTP-CCSD($\sfrac{1}{2}$) & 0.29 & \textbf{\emph{1.62}} & 0.48 & \textbf{\emph{1.97}} & -0.05\tabularnewline
\hline 
TP-CCSD($\sfrac{1}{4}$) & \textbf{\emph{0.74}} & -0.42 & \textbf{\emph{0.50}} & \textbf{\emph{1.21}} & \textbf{\emph{2.29}}\tabularnewline
\hline 
XTP-CCSD($\sfrac{1}{4}$) & \textbf{\emph{0.69}} & \textbf{\emph{-0.59}} & \textbf{\emph{0.55}} & \textbf{\emph{1.14}} & \textbf{\emph{2.34}}\tabularnewline
\hline 
\end{tabular}

}

\subfloat[Estimated excess kurtosis values for each distribution.]{%
\begin{tabular}{|c|>{\centering}p{0.15\textwidth}|>{\centering}p{0.15\textwidth}|>{\centering}p{0.15\textwidth}|>{\centering}p{0.15\textwidth}|>{\centering}p{0.15\textwidth}|}
\hline 
CVS-EOM-CCSD & -0.48 & \textbf{\emph{-0.87}} & \textbf{\emph{-1.19}} & -0.04 & \textbf{\emph{15.40}}\tabularnewline
\hline 
CVS-EOM-CCSD{*} & 0.59 & \textbf{\emph{1.02}} & -0.18 & -- & --\tabularnewline
\hline 
TP-CCSD($\sfrac{1}{2}$) & -0.33 & 0.42 & 0.56 & \textbf{\emph{3.20}} & \textbf{\emph{10.43}}\tabularnewline
\hline 
XTP-CCSD($\sfrac{1}{2}$) & -0.77 & \textbf{\emph{1.62}} & \textbf{\emph{-0.83}} & \textbf{\emph{3.28}} & \textbf{\emph{9.21}}\tabularnewline
\hline 
TP-CCSD($\sfrac{1}{4}$) & -0.27 & -0.42 & \textbf{\emph{-1.03}} & \textbf{\emph{0.91}} & \textbf{\emph{15.30}}\tabularnewline
\hline 
XTP-CCSD($\sfrac{1}{4}$) & -0.13 & -0.59 & -0.63 & 0.68 & \textbf{\emph{16.08}}\tabularnewline
\hline 
\end{tabular}}

\caption{\label{tab:Statistical-measures-measuring}Statistical measures measuring
deviations from a normal distribution. See Table \ref{tab:Normality-values-for}
for column headings. Significant deviations from normality are highlighted.}
\end{table}

In the preceding sections, it is assumed that the deviations in each
category (absolute/relative excitation energies, ionization potentials,
and absolute/relative oscillator strengths) represent a normal distribution.
In order to test this assumption, we have constructed normal quantile-quantile
(Q-Q) plots. In this approach, all observations (deviations) within
a category are sorted and then plotted against the inverse cumulative
normal distribution (ICDF) for a set of uniformly-distributed probability
values. For the \emph{i}th sorted observation out of $n$ (starting
with 1), the probability value $p_{i}=(i-0.5)/n$ is input to the
ICDF $Q(p)=\sqrt{2}\operatorname{erf}^{-1}(2p-1)$ to obtain the value
to plot against. For a true normal distribution, the Q-Q plot is a
straight line; thus, we perform a linear regression and report the
$R^{2}$ values as the ``normality'' scores. These values, in addition
to estimated skewness and excess kurtosis of the distributions are
given in \tabref{Statistical-measures-measuring}.

The distributions for absolute excitation energy deviations are all
highly normal, with at most a slight skew towards positive deviations.
Moving to relative excitation energies shows additional deviations
from normality. The most pronounced deviation is for XTP-CCSD($\sfrac{1}{2}$),
which exhibits a significant positive skew, but also a large excess
kurtosis, indicating a sharp distribution, but with a long tail of
outliers towards the positive end. The excess kurtosis for CVS-EOM-CCSD{*}
is also somewhat positive, while CVS-EOM-CCSD{*} has a fairly negative
excess kurtosis, indicating a more flattened distribution. Despite
a more tightly peaked distribution for XTP-CCSD($\sfrac{1}{2}$),
the long positive tail is concerning for routine application; a similar
effect is seen in the larger maximum errors for XTP-CCSD($\sfrac{1}{2}$)
compared to TP-CCSD($\sfrac{1}{2}$). The ionization potential distributions
are almost all flattened (with the exception of TP-CCSD($\sfrac{1}{2}$)),
with excess kurtosis ranging from -0.63 to -1.19 (essentially a uniform
distribution). The distributions are slightly skewed (positively,
except for CVS-EOM-CCSD{*}), but it seems clear that a normal distribution
is sufficient for capturing the overall improvement of the TP-CC methods.

The distributions for oscillator strength deviations are much more
non-normal. In particular, the relative oscillator strengths exhibit
extremely large excess kurtosis, suggesting much more sharply peaked
distributions that the standard deviation in \figref{rel-int} would
seem to indicate, but with long tails in both directions (more to
the positive for CVS-EOM-CCSD and (X)TP-CCSD($\sfrac{1}{4}$)). The
primary outliers seem to be the fluorine edge of $\ce{CH3F}$ as well
as the oxygen edge of $\ce{CH3OH}$. In the former case, the most
intense peak in most calculated spectra is a Rydberg $3p$ transition;
inclusion of additional diffuse basis functions would likely decrease
the intensity of this and the other Rydberg peaks and so should lead
to better overall accuracy. The absolute oscillator strengths are
somewhat more well-behaved, but still with significant tails to the
positive side.

\section{Conclusions and Future Work}

Core ionization and core excitation energies for a variety of small
molecules were calculated with several ``traditional'' equation-of-motion
coupled cluster methods: CVS-EOM-CCSD, -CCSD{*}, and -CCSDT, as well
as with a range of proposed transition-potential coupled cluster methods,
(X)TP-CCSD($\lambda$), in which a fractionally-ionized or -excited
reference state is employed. In comparison to full CVS-EOM-CCSDT,
TP-CCSD($\sfrac{1}{2}$) seems to account for essentially all of the
orbital relaxation energy. CVS-EOM-CCSD{*} was used as an ``aspirational
yardstick'' in this study, as previous work showed that a simple
non-iterative triples correction in the final state was able to account
for the relaxation error, and TP-CCSD($\sfrac{1}{2}$) indeed achieves
similar statistical error measures. The XTP-CC variant, in which the
fractional core electron is promoted to the LUMO, does not measurably
improve on TP-CC, despite expectations that a neutral reference state
should provide more balanced orbitals (especially virtual orbitals).
TP-CCSD($\sfrac{1}{4}$) was also tested, and has intermediate performance
between CCSD and TP-CCSD($\sfrac{1}{2}$), while TP-CCSD($\sfrac{3}{4}$)
and TP-CCSD(1) calculations were seen to diverge.

Overall, TP-CCSD($\sfrac{1}{2}$) is about as accurate for core-excited
states as EOM-CCSD is for valence states, with deviations from full
CVS-EOM-CCSDT within a few tenths of an eV. TP-CCSD intensities are
also much improved over CVS-EOM-CCSD, and in particular considering
normalized spectra the reduction in deviations is between three- and
four-fold. The ability of a relatively simple modification of ``vanilla''
CVS-EOM-CCSD, using only non-standard reference orbitals, to accurately
predict core ionization and core excitation energies is an exciting
prospect, especially as the computational cost of TP-CCSD is essentially
the same as CVS-EOM-CCSD and scales rigorously with system size $n$
as $\mathscr{O}(n^{6})$.

While this study looks at only a few choices of $\lambda$ (the fraction
of electron to remove from the core orbital), it would be interesting
to pursue optimizing $\lambda$, possibly in an element-specific manner.
In addition, B3LYP orbitals are used here, in order to capture some
effect of electron correlation on the fractionally-occupied reference
state. In future work, we plan to assess the use of plain Hartree--Fock
orbitals which would allow for a simpler native implementation in
CFOUR. Additionally, we are interested in applying TP-CCSD to the
case of transient x-ray absorption (tr-NEXAFS, alternatively UV pump/x-ray
probe or PP-NEXAFS); the ability of the TP orbitals to simultaneously
describe valence excitations is of highest importance. Lastly, the
application of TP-CCSD to larger molecules and direct comparison with
experimental spectra is necessary to fully explore the potential benefits
of our proposed method.
\begin{acknowledgments}
This work was supported by a generous start-up grant from Southern
Methodist University. All calculations were performed on the ManeFrame
II computing system at SMU. We would also like to thank Dr. Christopher
Ehlert for his work on PSIXAS which made this project possible.
\end{acknowledgments}

\section*{Data Availability}

The data that supports the findings of this study are available within
the article and its supplementary material.\bibliographystyle{unsrt}
\bibliography{paper}

\end{document}